\documentclass[amsmath,amssymb,pra,aps,showpacs,twocolumn,10pt,superscriptaddress]{revtex4-2}
\usepackage[colorlinks=true,citecolor=blue,linkcolor=red,urlcolor=blue]{hyperref}
\usepackage{graphicx}
\usepackage{dcolumn}
\usepackage{bm}
\usepackage[utf8]{inputenc}
\usepackage[T1]{fontenc}
\usepackage{lmodern}
\usepackage{cleveref}
\usepackage{color}
\usepackage{amsmath}
\usepackage{amssymb}
\usepackage{amsfonts}
\usepackage{latexsym}

\crefname{equation}{}{}

\begin{document}

\title{Simple Hamiltonian dynamics is a powerful  resource for image classification}

\author{Akitada Sakurai} \email{akitada.sakurai@oist.jp}
\affiliation{Okinawa Institute of Science and Technology Graduate University, Onna-son, Okinawa 904-0495, Japan}
\author{Aoi Hayashi}  
\affiliation{School of Multidisciplinary Science, Department of Informatics, SOKENDAI (the Graduate University for Advanced Studies), 2-1-2 Hitotsubashi, Chiyoda-ku, Tokyo 101-8430, Japan}
\affiliation{Okinawa Institute of Science and Technology Graduate University, Onna-son, Okinawa 904-0495, Japan}
\affiliation{National Institute of Informatics, 2-1-2 Hitotsubashi, Chiyoda-ku, Tokyo 101-8430, Japan}
 \author{W. J. Munro}
\affiliation{Okinawa Institute of Science and Technology Graduate University, Onna-son, Okinawa 904-0495, Japan}
 \affiliation{National Institute of Informatics, 2-1-2 Hitotsubashi, Chiyoda-ku, Tokyo 101-8430, Japan}
\author{Kae Nemoto}\email{kae.nemoto@oist.jp}
\affiliation{Okinawa Institute of Science and Technology Graduate University, Onna-son, Okinawa 904-0495, Japan}
\affiliation{National Institute of Informatics, 2-1-2 Hitotsubashi, Chiyoda-ku, Tokyo 101-8430, Japan}

\date{\today}

\begin{abstract}
A quadrillion-dimensional Hilbert space hosted by a quantum processor with over 50 physical qubits has been expected to be powerful enough to perform computational tasks ranging from simulations of many-body physics to complex financial modeling.  Despite a few examples and demonstrations, it is still unclear how we can utilize such a large Hilbert space as a computational resource, particularly how a simple and small quantum system could solve non-trivial computational tasks. This paper shows a simple Ising model capable of performing such non-trivial computational tasks in a quantum neural network model.   An Ising spin chain as small as ten qubits can solve a practical image classification task with high accuracy.  To evaluate the mechanism of its computation, we examine how the symmetries of the Hamiltonian would affect its computational power.   We show how the interplay between complexity and symmetries of the quantum system dictates the performance as quantum neural network.

\end{abstract}
\maketitle

\section{Introduction}
Recent developments in the field of quantum computation have seen processors composed of hundreds of qubits realized which can be used to simulate novel phenomena for a variety of physical, chemical, and biological systems, as well as solving interesting financial and mathematical problems~\cite{Lanyon2010, Aaronson2013, Rebentrost2018, Nam2020, Carrera2021, Xu2021, Gong2022}. These processors are large enough that they should be able to generate sufficiently complex quantum dynamics so that they can be used to simulate and explore the nature of quantum many-body physics~\cite{Britton2012, Jurcevic2014, Guo2021, Ebadi2021, Ippoliti2021, Thanasilp2021, Philipp2022}. If such an expectation were feasible, then we could consider whether it would be possible to reverse this process and use a typical many-body system, such as the Ising model, to perform computational tasks?  This leads to a very interesting question: would natural quantum dynamics alone be a useful resource for quantum computation?  If so, would the common beliefs for the requirements for quantum computation change, such as the necessity of non-integrable systems for quantum machine learning~\cite{Xia2022} and circular unitary ensemble (CUE) processes for the quantum reservoir~\cite{Hayashi2022,Kubota2023}?  Do the commonly used measures for computational capacity, such as the information processing capacity~\cite{Dambre2012, Martinez2023, GarciaBeni2023} remain appropriate?  In this article, we aim to answer these questions by designing a quantum computational model using only the Ising model for its quantum contribution.    

\begin{figure}[htb]
\centering
\includegraphics[width=0.47\textwidth]{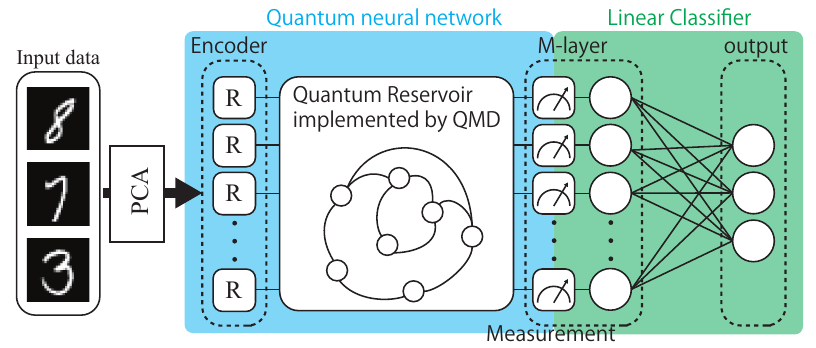}
 \caption{Schematic illustration of the structure for quantum extreme reservoir computation (QERC). Here, the blue and green areas represent the quantum and classical components, respectively. At the first layer, inputs (vectors) are transformed into vectors in a $2N$-dimensional space using principal component analysis (PCA). These components are then encoded into our quantum system using single-qubit rotation gates. Quantum many-body dynamics (QMD), such as the Ising models, map the input to complex final states. The measurement obtains probability distributions and is then transformed to a distribution with zero mean and one standard deviation. We refer to the measurement and this transfer together as the M-layer.  Then, the distribution is sent to a linear classifier. The training process is only applied to the classical neural network, while the quantum reservoir implemented using QMD is fixed.}
\label{QERC}
\end{figure}

To explore and address these questions, we benchmark quantum neural networks using quantum extreme reservoir computation (QERC)~\cite{Akitada2022}.  This QERC process, as depicted in Fig. \ref{QERC}, is an extreme machine learning model with a quantum reservoir for the middle hidden layer of nodes.  As such, the quantum reservoir is the only quantum component in the system. The information in the quantum reservoir would be extracted and converted to classical data during the measurement step. It is then fed to a one-layer classical neural network, a linear classifier for classification problems.  The quantum reservoir provides a given quantum dynamics, while the total system works as a quantum neural network model with the dual nature associated with extreme machine learning and reservoir computation. More specifically, this is a two-layer neural network with a large node number and no optimization nor control on the recurrent neural network (the reservoir).

Among the various quantum machine learning models, we choose the QERC for this benchmark for several reasons.  First and foremost, variational quantum algorithms~\cite{Peruzzo2014,Mitarai2018, Magann2021}, which are a class of quantum neural networks, have been extensively studied for quantum phase transitions and determining quantum eigenstates \cite{Colless2018,Uvarov2020,Gong2022}. However, these models require the ability to design a quantum circuit with a parameter that is optimized by classical computation. That optimization is a known hard optimization problem in these models~\cite{Larocca2022} . Further, it is generally difficult to implement the quantum circuit needed to realize the required feature space~\cite{Havlicek2019,Abbas2021}. Second, quantum reservoir computation~\cite{Fujii2017,Nakajima2019, Chen2020, Angelatos2021, Suzuki2022, Bravo2022, Domingo2022} could be closer to the QERC model. However, it was constructed as a quantum extension of classical reservoir computation and designed to analyze temporal data.  Quantum extreme learning machine had also been conceptually proposed as an extension of the two-layer neural network to be replaced by a quantum neural network~\cite{Mitarai2018, fujii2020}.  Although, unlike the concept of the quantum perceptron~\cite{Altaisky2001, Sanjay2001, panella2011}, these models could be consistent with quantum mechanics, they are not suitable as a benchmarking candidate until they show at least one concrete instance where they can demonstrate their capability to achieve a significantly high accuracy.    

By contrast, the QERC model has an explicit form of computation and requires only the initialization of the quantum reservoir to perform various tasks.  The (not only conceptual but also) explicit form of the QERC is essential to evaluate the validity of the measures and requirements commonly used for quantum computational capacity.  Moreover, it has numerically demonstrated the accuracy necessary for benchmarking with few qubits. This is essential for the numerical evaluation.  Now, the estimation of the quantum computational power of quantum machine learning models tends to focus on the complexity of the quantum process.  However, a quantum neural network model provides its capability through the process from the classical input to the classical output. Hence, the complexity of the quantum process is not enough to evaluate the capability of the quantum neural network.  In the field of computer science, benchmarking has long been used for the evaluation of computational capability.

In this paper, we employ benchmarking of the quantum reservoirs within the QERC model to evaluate their requirements and capacity measures. To achieve this, we must first fix the nature of the output measurements.  As such measurements can be as computationally powerful operations, similar to the universality of gate-based quantum computation~\cite{gottesman1998}, our measurements in this QERC model will be a projective measurements on to the computational basis. These can be implemented using single-qubit $Z$-axis measurements.  Next, the system size in our benchmarking will be around 10 qubits. As such, we need to choose an appropriate task to run our benchmarking on, and the MNIST data set for hand-written numbers~\cite{lecun2010} has sufficient complexity for this size regime.

\section{Quantum reservoir models}
It is well known that any quantum operation can be decomposed with an arbitrary accuracy with a universal gate set involving only single and two-qubit operations \cite{aharonov2003,Sawicki2017}. It has been proven that multi-qubit operations are unnecessary \cite{aharonov2003, Sawicki2017}.  However, in Hamiltonian-based computation, its complexity depends on the types of many-body interactions involved~\cite{Biamonte2008, Kempe2003}. For example, a system with classical many-body interactions and a transverse magnetic field, such as the Ising model used in adiabatic quantum computation, shows a different complexity between the two-body situation and those above~\cite{Lucas2014,Kempe2003}. The previous work~\cite{Akitada2022} used a discrete time crystal model consisting of a time-periodic Hamiltonian with only two-body interactions, but its effective Hamiltonian has more than three-body interactions. It is not, therefore, clear that a two-body Hamiltonian is enough to achieve the same performance in QERC. To evaluate whether two-body Hamiltonians are sufficient, let us first employ two Ising models that consist only of one- and two-body interactions: the ZZ-Ising model with the corresponding Hamiltonian,
\begin{equation}
\hat{H}_{ZZ} = \hbar \sum_{l>m}J_{lm}\sigma_l^z\sigma_m^z + \hbar  g \sum_l\sigma_l^x,
\label{eq: zz ising}
\end{equation}
and the XX-Ising model given by the Hamiltonian
\begin{equation}
\hat{H}_{XX} = \hbar \sum_{l>m} J_{lm}\sigma_l^x\sigma_m^x +   \hbar  g \sum_l\sigma_l^z,
\label{eq: xx ising}
\end{equation}
where $\sigma_l^\mu$ $(\mu=x,y,z)$ are the usual spin-1/2 Pauli matrices at the $l$-th site.  Next $J_{lm}$ and $g \geq 0$  are the strength of the two-qubit interaction and the transverse magnetic field, respectively.  We set $J_{lm} = J_0/|l-m|^\alpha$, where $J_0$ is the strength of the interaction and $\alpha$ is a long-range order parameter.  These Hamiltonians have been experimentally demonstrated in a variety of different physical platforms, including~\cite{Porras2004, Britton2012, Labuhn2016, Zhang2017, Zhang2017Ising, Choi2017,Bluvstein2021, Mi2022}.  In those trapped ions demonstrations ~\cite{Porras2004, Zhang2017, Zhang2017Ising} $\alpha\sim1.5$..

Now the XX-Ising and ZZ-Ising models are related by the unitary transformation $\hat{H}_{XX} = \exp{(i\sum_l\frac{\pi}{4}\sigma_l^y)} \hat{H}_{ZZ} \exp{(-i\sum_l\frac{\pi}{4}\sigma_l^y)}$. Both Hamiltonians have a U(1) symmetry; however, only the ZZ-interaction shares the symmetry with the computational basis measurement.  This difference can affect the performance of a quantum reservoir, as we will observe later.

 \begin{figure*}[htb]
\centering
\includegraphics[width=0.8\textwidth]{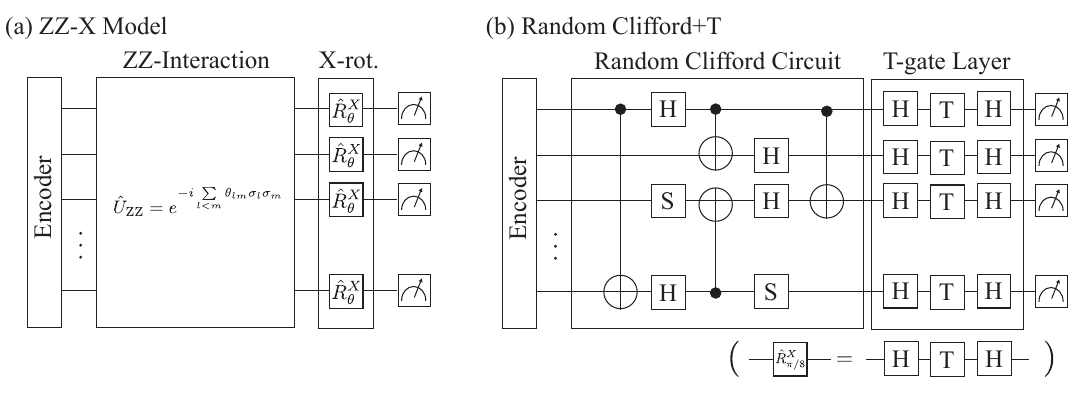}
 \caption{Depiction of the quantum computational circuit for the ZZ-X and Cliffort+T models. In (a) for the ZZ-X model, the ZZ-interaction part can be decomposed into a circuit of quantum gates. However, it can be advantageous for the ZZ-X model to have the ZZ-interaction part created using Hamiltonian dynamics, which would not require any gate-circuit decomposition. As such, the model's simplicity allows one to use more specialized quantum computational devices such as quantum annealer.  Next (b) depicts the Clifford+T model, where the Clifford circuit consists of only randomly selected Clifford gates, followed by one layer of T-gates before measurement.}
\label{ZZXandClifordT}
\end{figure*}

Next, given the XX-Ising and ZZ-Ising models, let us consider two further models.  The first is the ZZ-X model whose dynamics are governed by
\begin{equation}
\hat{U}_\mathrm{ZZX} = \exp\left(-i\sum_{l}\theta^x_l\sigma_l^x\right)\exp\left(-i\sum_{l<m}\theta^{J}_{lm}\sigma_l^z\sigma_m^z\right)
\label{eq: zzx unitary}
\end{equation}
where $\theta_l^x$ and $\theta_{lm}^J$ are rotation angles.  To simplify our analysis further, we set $\theta_l^x=\theta_x$ to be equal over the site, similar to the transverse field of the Ising model. Now when $\pi/2\leq \theta_x< \pi$, the rotation part can be written as,
\begin{equation}
\exp\left(-i\sum_{l}\theta^x_l\sigma_l^x\right) = \hat{P}_x \exp\left(-i(\theta_x-\pi/2)\sum_{l}\sigma_l^x\right),
\end{equation} 
where $\hat{P}_x$ is the parity operator. Because $\hat{P}_x$ is a bit-flip operation and the computational basis measurement is followed, the permutation can be absorbed into a linear transformation within the linear classifier. Thus, we investigate only the $0\leq \theta_x \leq \pi/2$ range in the analysis. Next, the angle $\theta_{lm}^{Jt}$ is site-dependent, and like for the Ising model, we consider the long-range interaction to be represented by $\theta^{J}_{lm} = \theta_J/|l-m|^\alpha$ with $\theta_J$ being the maximum angle.   

To be more computational complexity oriented, our second additional model is associated with a quantum circuit of the random Clifford gate circuit plus one depth of T-gates for each qubit as the quantum reservoir.  Fig.~\ref{ZZXandClifordT} depicts our various strategies where the ZZ-X model has two circuits: the natural dynamics of the ZZ-Ising model and the single X rotation on each qubit as seen in Fig.~\ref{ZZXandClifordT} (a).  Next, Fig.~\ref{ZZXandClifordT} (b) shows the fully gate-decomposed model with a random Clifford circuit and the T-gate layer. This is referred to as the Random Cilifford+T model in the remainder of this paper. 

Finally, we employ Haar measure sampling of the unitary matrices as the standard model to evaluate others against.  To evaluate this non-trivial dynamics of quantum computation, randomness is often used as an estimator for the computational capability in both experimental demonstrations~\cite{Arute2019,Thanasilp2021} and quantum variational algorithms~\cite{Abbas2021,Menon2021}.  Haar measure sampling has also been used to design quantum machine learning models~\cite{Hayashi2022}, which are deeply connected to information scrambling in the computational process. 

  Haar measure random sampling may be ideal for information scrambling. However, it is not necessarily easy to extract information from such a final state. In the process of quantum computation, we expect to utilize the Hilbert space to gain quantum advantage, and the process for the input state to be spread in the Hilbert space is called information scrambling.  Then at the stage of the readout, ideally the solution state is to be one of the computational basis, which is then read out by a single-shot computational basis measurement.  Hence, in general we have two factors in quantum computation: information scrambling and conversion to/towards the solution.  The complication of quantum-classical hybrid models is that we do not know which part of the model carries what role, or even if we define each role to each part of the model.  Generally, quantum machine learning models including QERC face this complication.  For instance, usually, the circuit before the variational model in VQA is considered for encoding and information scrambling, however, that is only a convenient recognition to design the model.  Similarly, in QERC, it is convenient to think that the quantum reservoir is there to exploit the large Hilbert space, indicating information scrambling. However, there is no necessity for the model to divide the task in this way.  Hence, it is necessary to include Haar measure random sampling in the benchmark to understand the mechanism of the computation. We choose a particular Haar-random unitary in this case.

\section{Benchmarking}
In this benchmarking, we will evaluate the quantum reservoir models described in the previous section as they are used in the QERC shown in Fig.~\ref{QERC}.   It is important to remember that in the QERC model, there are two extra network layers: the encoding of classical data to the quantum reservoir and the measurement of the quantum reservoir.  As we use the MNIST hand-written digits (0 to 9) data set for our benchmark, we need to use the principal component analysis (PCA) to encode the $2N$ largest principal components to $N$ spins by single rotations on the initialized state $|0\rangle\otimes\cdots\otimes|0\rangle$ as 
\begin{equation}
|\psi_l\rangle = \cos\left(\theta_l/2\right)|0\rangle +  e^{i\phi_l}\sin\left(\theta_l/2\right)|1\rangle
\label{eq:encode}
\end{equation}
with $0\leq \theta_l,\phi_l\leq \pi$~\cite{Akitada2022}. 
For each image, the first $N$ elements are encoded on the angles $\theta_l$, while the next $N$ components are encoded on the phase $\phi_l$. The smaller components are ignored.  This paper considers $10$ sites for the Ising model with $10$ spins as qubits.   

The quantum reservoir readout is achieved by single-qubit projective measurements on the computational basis at the M-layer~\cite{Akitada2022}.  The output classical data is then treated and processed by the one-layer classical neural network performing the gradient
descent AdaGrad~\cite{duchi2011Adaptive}  machine learning algorithm as shown in Fig~\ref{QERC} (see appendix~\ref{AppendixQERC} for details on the classical neural network).  It is important to note that the M-layer is a middle layer in the neural network of the QERC, and the measurement takes a computational role.  In this paper, the computational basis states 0 and 1 are the eigenstates of the $Z$ operator for each qubit.

\section{Performance} \label{performance}

\begin{figure}[htb]
\centering
\includegraphics[width=0.4\textwidth]{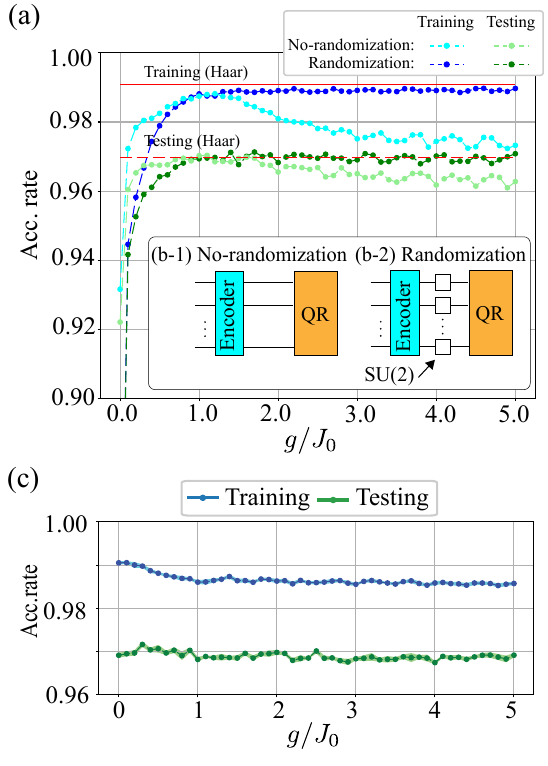}
 \caption{(Plot of the $g$-dependence of the ZZ-Ising/XX-Ising models for a scaled time $J_0 t = 3.5$. (a) shows the $g$-dependence of the ZZ-Ising model with ten sites, where the transverse magnetic field is applied along the x-axis. The blue and green lines show the training and testing results for the encoder in the inset (b-1), while the cyan and light-green lines show the training and testing results for the encoder with random single qubit rotation shown in (b-2). (c) illustrates the result of the g-dependence with $\alpha=1.5$ of the XX-Ising model. The points and shadows are the mean value and the standard deviation taken from 40 to 50 epochs to reveal the stability of the optimization.} 
\label{ZZ-Ising}
\end{figure}

Let us begin by taking the ZZ-Ising model as the quantum reservoir of the QERC to evaluate its performance.  Fig.~\ref{ZZ-Ising} shows the results for the two different strategies with this quantum reservoir.  The light blue dots (training) and the light green dots (testing) are for the ZZ Ising model while the blue dots (training) and the green dots (testing) are for the ZZ Ising model with the randomization of single qubit rotation before the time-evolution of the quantum reservoir as shown in the inset (b-2).  We have allowed the system to evolve for a scaled time $J_0 t = 3.5$.  

Although the training accuracy for the Haar measure sampling (indicated as the solid red line in Fig.~\ref{ZZ-Ising}) is slightly better than the highest accuracy with the ZZ Ising model, the highest accuracy with the ZZ-Ising model at $g/J_0\sim 1$ achieves the accuracy as high as the testing accuracy given by Haar measure random sampling.  The degradation of accuracy in the strong transversal magnetic field regime $(g/J_0>1)$ can be overcome by the randomization of the single qubit rotation.  It is worth noting that the single-qubit gates alone (without the quantum reservoir) reduce the accuracy rate in this model.  This indicates that this process does not directly contribute to the computation but converts the input state in a form that the quantum reservoir can be more effective. Although the transversal magnetic field is necessary for the QERC to perform, the strong transversal magnetic field introduces the bias in the quantum dynamics, which may cause a mismatch between the tendency in the input data and the quantum reservoir similar to the behavior observed in classical machine learning ~\cite{haykin2009}. 

Next, we explore the QERC's performance using the XX-Ising model given in Eq. ~\ref{eq: xx ising}. We begin by showing in Fig.~\ref{ZZ-Ising} (c) the g-dependence of the XX-Ising model, where we observe that its performance is high even in the $g/J_0<1$ regime. That shows that the XX-Ising model can utilize the many-body interaction without the transverse field for computation.

This result indicates that only the two-body interaction is enough to generate a powerful reservoir with high explanation power. However, since ZZ-Interaction has U(1)-symmetry along the z-axis, which is the same as our measurement axis, its power does not appear when $g=0$ in the ZZ-Ising mode. In summary, in QERC, when generating strong quantum reservoirs from quantum dynamics, complex interactions and manipulations such as three-body and driven interactions are essentially unnecessary, and two-body interactions are sufficient. While, as in the case of ZZ-Ising, the system's symmetry may mask the interactions' ability, and additional manipulations or interactions are needed to eliminate or reduce the effect of symmetry.

\begin{figure*}[htb]
\centering
\includegraphics[width=0.98\textwidth]{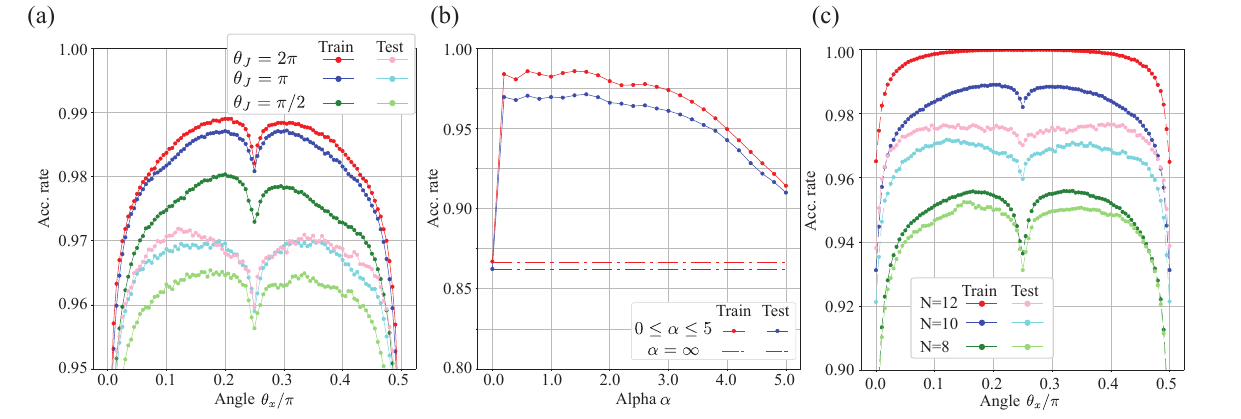}
\caption{Classification performance of the ZZ-X model. In (a) we plot the performance with angles $\theta_J = \pi/2, \pi, 2\pi$ and $0\leq \theta_x \leq 0.5\pi$ for $N=10$ sites. Then in (b) and (c) we show the $\alpha$-dependence and size-dependence of the ZZ-X model, respectively. Here we have set $\theta_J = 2\pi$ with $\theta_x = 0.125\pi$ for both plots. The stability of the optimization is shown with shadows to present the standard deviations obtained from 40 to 50 epochs.}
\label{ZZX}
\end{figure*}

Now, let us explore our two models that have a quantum circuit structure.  Our first evaluation is for the ZZ-X model given by Eq.~\ref{eq: zzx unitary}, noting that this model is exactly the same as the ZZ-Ising model without the transversal field.  It is interesting to see if the single qubit rotation is enough to extract the potential complexity generated through the ZZ-Ising interaction.  Fig.~\ref{ZZX} (a) illustrates the performance of the ZZ-X model dependency on the angle $\theta_x$ for each different angle $\theta_J$, while (b) shows the $\alpha$ dependency on the performance. Next (c) shows the performance dependency on the system size for $\theta_J=2\pi$.  The dip in the accuracy rate around $\theta_x=\pi/4$ in Fig.~\ref{ZZX} (a) for $N=10$ is smaller for a larger number of qubits.  As the number of qubits increases, the wider range of $\theta_x, \alpha$ allows us to achieve a high accuracy performance for our classification task.
\begin{table}[htb]
\begin{tabular}{|c|l|l|}
\hline
model      & \multicolumn{1}{c|}{Train acc. (std.)} & \multicolumn{1}{c|}{Test acc. (std.)} \\ \hline
Haar &     $0.990\pm 0.0004$          &       $0.969\pm 0.0004$                          \\ \hline
SU(2) random circuit&  $0.812\pm 0.0011$          &       $0.811\pm 0.0015$                          \\ \hline
ZZ-Ising  &     $0.987\pm 0.0004$          &       $0.970\pm 0.0006$                          \\ \hline
XX-Ising  &   $0.986\pm 0.0004$            &        $0.968\pm 0.0006$                         \\ \hline
ZZ-X       &       $0.986\pm 0.0003$           &           $0.972\pm 0.0005$                      \\ \hline
T gates (HTH)    & $0.867\pm 0.001$                       & $0.8627\pm 0.001$              \\ \hline
Random Clifford   & $0.902 \pm 0.001$                      & $0.884\pm0.002$                \\ \hline
Random Clifford+T & $ 0.990\pm 0.000$                      & $0.970\pm0.001$                \\ \hline\hline
Linear Classifier& $ 0.936\pm  0.00029$                      & $0.928\pm0.0007$                \\ \hline
Linear Classifier (20 pc)& $ 0.874\pm 0.00025$           & $0.879\pm0.0007$                \\ \hline
\end{tabular}
\caption{Training and testing reservoir accuracy rates for models. Here, as the linear classifier, we used a single-perceptron with the softmax activation function. The last one is the result of the linear classifier with PCA (20 pc).} 
\label{performance-table}
\end{table}

Our second quantum circuit structure model is the Clifford+T model and we present its performance in Table~\ref{performance-table} along with the other models we have considered. We clearly observe that the Clifford+T model has the best performance in terms of both the training and testing accuracy. The ZZ-Ising, XX-Ising, and ZZ-X models are, however, within 0.5\%.

\section{Entanglement and Integrability}
Given the above performance benchmarking using the various quantum reservoir models, we now turn our attention to the properties of these quantum reservoirs.  Entanglement is known to be necessary for universal quantum computation and is often discussed in quantum machine learning~\cite{Pfeffer2022}.  In the benchmark, we observe that the quantum reservoirs that have achieved a higher accuracy than the classical linear classifier provide generate entanglement in the computational process.  By contrast, the quantum reservoir of random single-qubit rotations gives performance poorer than the classical linear classifier.   This implies that entanglement is a key resource for the performance of QERC.  Within the numerical observation, QERC is not an exception to quantum algorithms in general regarding the role of entanglement~\cite{jozsa2003}.   It is also known that entanglement does not guarantee the computational power that quantum computation promises.  This is also shown in our benchmark, which we will explore next.

The ZZ-Ising model with $g/J_0=0$ does not give any improvement, although it generates entanglement in the output state.  This arises simply from the symmetry shared by both the dynamics and the measurement.  The measurement used at the M-layer does not access to the effect of the entanglement generated in the state.  Hence, the comparison of this model with the ZZ-X model gives us an interesting insight.  The difference between the ZZ-Ising model with  $g/J_0=0$ and the ZZ-X model is the last layer of single-qubit gates in the circuit for the ZZ-X model, yet the ZZ-X model can achieve as high as any other quantum reservoir models.  Fig.~\ref{ZZX} (a) shows the performance (accuracy rate against the single qubit rotation $\theta_x$.  The optimal points appear around the angle  $\pi/8$, which is equivalent to the T-gate. As we know, single-qubit rotations alone cannot contribute to the performance; this indicates that changing the measurement axis is sufficient to extract the entangling dynamics of the ZZ-Ising interaction created for the computation. This suggests that the complexity the ZZ-Ising model can generate is sufficient for high QERC performance, which agrees with the XX-Ising model performance, noting that the XX and ZZ interactions are related by a simple unitary transformation. A more interesting comparison can be seen with the Clifford circuit and the Clifford +T model.  It is known that a Clifford circuit can be efficiently simulated in a classical computation \cite{Aaronson2004,nest2009} despite of the entanglement in the computational system.  Our numerical simulation shows that the single layer of T gates increases the accuracy rate.  The role of the T-gates is similar to the computational complexity in learning a probability distribution, where a single T-gate was used~\cite{Hinsche2023}.  In our case, a single T gate is insufficient to recover the accuracy rate fully.

\begin{figure}[htb]
\centering
\includegraphics[width=0.45\textwidth]{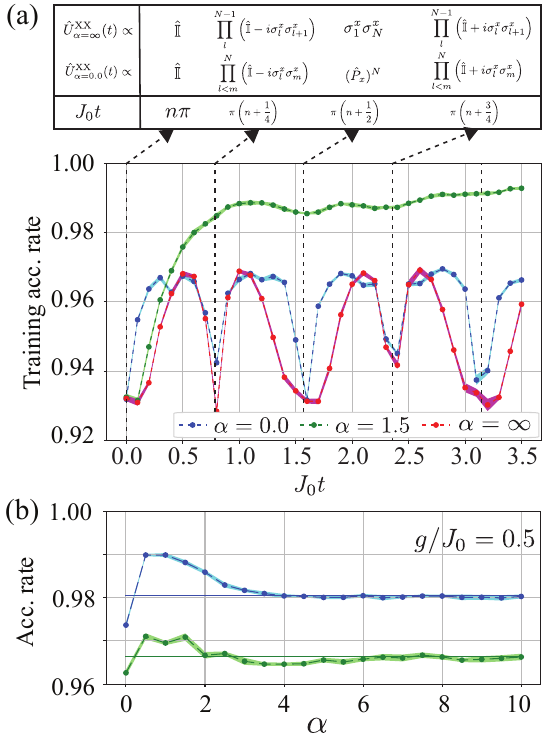}
\caption{Plot of the training accuracy rate of XX-Ising model for three different values of $\alpha= 0.0$(blue), $1.5$(green), $\infty$(red). Here we have set $g/J_0 = 0.0$. The shadows displaced show the standard deviations obtained from 40 to 50 epochs. The inset table shows the unitary operators at the stroboscopic times, $J_0t = n\pi, (n+1/4)\pi,  (n+1/2)\pi$ and $(n+1/3)\pi$ with $n\in \mathbb N_{\ge 0}$ and where $N$ is the number of sites and $\hat{P}_X$ the parity operator along the x-axis. The points and shadows are the mean value and the standard deviation taken from 40 to 50 epochs to reveal the stability of the optimization. (b) shows the $\alpha$-dependence with $g/J_0=0.5$. The blue and green dashed lines with dots plot the training and testing accuracy with $\alpha$ ranging from $\alpha=0.0$ to $10$ (the blue and green solid lines indicate the training and testing accuracy with $\alpha=\infty$). }
\label{XX-Ising Alpha}
\end{figure}

Next, we look into the influence of the integrability of the XX-Ising model on its performance as a quantum reservoir.  Here, we consider the analytical form of the unitary map of the quantum reservoir and refer to its integrability. In the benchmark, we set $\alpha$ to be 1.5, as that parameter value gives the optimal performance.  To consider integrability in this case, we take different values of $\alpha$ to determine how its integrability affects the performance.  We will consider three different interaction ranges ($\ \alpha=0.0, 1.5$, and $\infty$). The first $\alpha=0$ corresponds to all-to-all connectivity with equal strength, whereas $\alpha=\infty$ corresponds to nearest neighbour coupling only. Now $\alpha=1.5$ sits between these two extremes and corresponds to the long-range order parameter in recent ion-trap experiments \cite{Porras2004, Britton2012, Zhang2017,Zhang2017Ising}.  The training accuracy rates are summarized for these three parameter values in Fig.~\ref{XX-Ising Alpha}.  
The best accuracy can be achieved for a middle range $\alpha$ value.  Unlike the case of $\alpha=1.5$, the $\alpha=0$ and $\infty$ cases show periodical behavior.  This reflects the periodic dynamics of the Hamiltonian with these particular values. It is useful to explore such mechanics more analytically.  Our unitary operator can be decomposed as  
\begin{equation}
\begin{split}
\hat{U}^{(XX)}_{\alpha}(t) &= \exp\left(-i\sum_{l>m}^{N}  \frac{J_0t}{|l-m|^{\alpha}} \sigma_l^x\sigma_{m}^x  \right)\\
&=  \prod_{l>m}^{N} \left( \cos\left(\omega_{lm} t\right)\hat{\mathbb{I}} -i \sin\left(\omega_{lm}t \right)\sigma_l^x\sigma_{m}^x \right),
\end{split}
\label{eq:XXIsing_alpha_infty}
\end{equation}
where the XX-interactions between each site commute with one another. In this case, the total unitary can be decomposed as a product of the simple oscillators with the angular frequencies $\omega_{lm}={J_0}/{|l-m|^{\alpha}}$. Thus, when $\alpha =1.5$, the oscillators are in different angular frequencies. This means the recurrence of dynamics is avoided, resulting in non-periodical dynamics.

Now, in the nearest-neighbour interaction case ($\alpha = \infty$), the oscillators may share the same angular frequency, introducing a symmetry. Our unitary operator may then be written as
\begin{equation}
\begin{split}
\hat{U}^{(XX)}_{\alpha=\infty}(t) &=  \prod_{l}^{N-1} \left( \cos\left(J_0t\right)\hat{\mathbb{I}} -i \sin\left(J_0t\right)\sigma_l^x\sigma_{l+1}^x \right)
\end{split}
\label{eq:XXIsing_alpha_infty}
\end{equation}
where we observe that the dynamics has a simple analytic form at times $J_0t= n\pi, (n+1/4)\pi, (n+1/2)\pi, (n+3/4)\pi$ (see the inset table in Fig.~\ref{XX-Ising Alpha}).This indicates that the unitary starts to grow its complexity over a short period of time, however at some point it starts to saturate and at $J_0t=n\pi$ its unitary returns to the identity. Consequently, the periodic nature does not allow the system to evolve for times long enough to perform our computation.

It is interesting also to explore the opposite limit of $\alpha=0$. Here, the Ising model also recovers the system symmetry, and the dynamics becomes solvable as
\begin{equation}
\begin{split}
\hat{U}^{(XX)}_{\alpha=0}(t) &=  \prod_{l<m} \left( \cos\left(J_0t\right)\hat{\mathbb{I}} -i \sin\left(J_0t\right)\sigma_l^x\sigma_m^x \right)
\end{split}
\label{eq:XXIsing_alpha_0}
\end{equation}
with all oscillators having the same angular frequency $\omega_{lm}=J_0$. Thus, similar to $\alpha=\infty$, the dynamics is analytic at the times $J_0 t= n\pi, (n+1/4)\pi, (n+1/2)\pi, (n+3/4)\pi$, where  its performance reduces, as shown in Fig.~\ref{XX-Ising Alpha}. When the dynamics is periodical, unless the period is extremely long, it is likely that the dynamics can be fully supported by a small subspace of the Hilbert space.  As shown in Fig.~\ref{XX-Ising Alpha}, the performance quickly recovers as $\alpha$ increases from zero, peaks around  $\alpha=1$, and gradually decreases to the large $\alpha$ limit. This indicates that the non-integrability of the quantum system contributes to the performance of the quantum neural networks.  
Interestingly, the power to generate entanglement is not sufficient to estimate the QERC performance as we have seen in above with $\hat{U}^{(XX)}_{\alpha=\infty}$ at $J_0 t = (n+1/4)\pi$ for instance.

This $\alpha$-dependence can be mitigated by applying a magnetic field in the $Z$-axis direction. In the ZZIsing model, the additional magnetic field in the $x$-axis direction breaks the U(1) symmetry. In contrast, in the XXIsing model, the additional field is necessary to resolve the integrability. Fig.~\ref{XX-Ising Alpha} (b) shows the $\alpha$-dependency of the accuracy rate.  The accuracy rate drops when $\alpha$ is large.  The value of the transversal field strength is chosen to minimize the $\alpha$-dependency, which is $g/J_0=0.5$.

Integrability impacts the performance of the QERC to some degree. However, it does not give us clear criteria for the quantum reservoir.  The ZZ-X model is an integrable system, though it does not hold U(1) symmetry due to its time evolution, which consists of two Hamiltonian components.  This indicates that when the time-evolution has a ``gate'' structure the integrability may not work as an indicator.  This is similar to the case of the Clifford gates $+$ T-gates.  The Clifford gate circuit can be efficiently simulated, and does not give us high performance as a quantum reservoir. However, the one-depth of T-gates can be enough to achieve similar high performance to the Haar measure sampling.  It may not be surprising that the integrability of the Hamiltonian system can significantly affect the performance of a quantum neural network.  However, the integrability is much less important in the ZZ-X model, which is {\it integrable} with the analytical form for the output state.
Integrability and periodicity are somewhat related in classical dynamics, and as we have seen above, the periodicity in the Hamiltonian dynamics prevents the states from evolving sufficiently for the one-layer classical neural network to distinguish their distributions. Non-integrability and non-periodicity of the system dynamics are not a necessary condition for a good quantum reservoir. However, they can still serve as an indication to provide good neural networks, as well as Haar random unitary does, which can be useful in designing quantum reservoirs.

\section{Information scrambling and sampling cost} \label{sampling}

In the previous section, we analyzed the properties of the quantum reservoirs with respect to their complex dynamics.  Here, we focus on the other element of quantum computation that can be achieved by quantum reservoir, that is, the convergence to the solution through the quantum algorithm.  So far, the quantum reservoirs we investigated require a linear classifier to complete classification tasks.  If a quantum reservoir can converge the state well enough to make the linear classifier redundant, such a quantum reservoir would be ideal.  The optimization of a quantum reservoir might be possible if we allow the quantum reservoir to be optimized.    Our question here is whether such a computational mechanism can be achieved without optimization of the quantum reservoir.   

To investigate this possibility, we look into the information scrambling and sampling cost for each quantum reservoir. The quantum dynamics provides a feature space for this computational model, and hence the dynamics complexity is important to deliver information scrambling necessary for the task. 

A unitary matrix sampled on the Haar measure should give information scrambling in our benchmark.   However, this process needs to be followed by the above computational mechanism, which is likely to be a clustering of states for classification problems.  It is unclear if a quantum reservoir has any contribution to this process.  If so, the quantum reservoir would help to work the linear classifier more efficiently.  To investigate this, we need to estimate the sampling cost for the measurement.

 \begin{figure}[htb]
\centering
\includegraphics[width=0.45\textwidth]{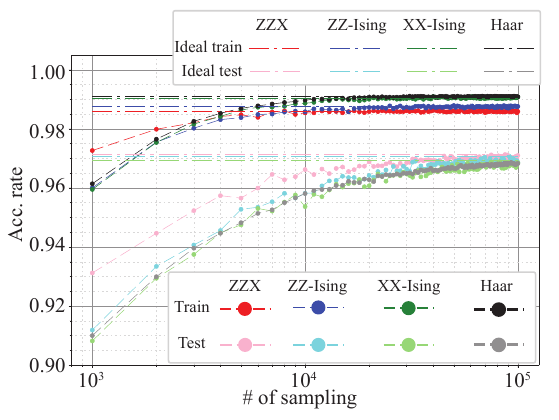}
 \caption{
 Sampling cost of QERC using different quantum reservoirs. We plot the sampling number dependence on the performance with the reservoirs (ZZ-Ising, XX-Ising, the Haar random and ZZ-X model), where we set $g/J_0 =1.0$ for the ZZ-Ising and $g/J_0=0.0$ for the XX-Ising model. In both models, we set $J_0t = 3.5$. The color lines show the accuracy rate for the probability distribution by assuming one can achieve the ideal situation. } 
\label{Sampling}
\end{figure}

The amplitude distribution of the final state of the quantum reservoir was used as the output of the M-layer in our benchmark, however in the simulations, we used the amplitude distribution reconstructed from the data.  Fig.~\ref{Sampling} shows the behavior of the accuracy rates dependent on the sampling number.   All the models in this comparison achieve similar accuracy rates with the theoretical amplitude distribution when the sampling number is large enough.  It has often been claimed that the necessary sampling number grows exponentially with the number of qubits for quantum algorithms that require the probability amplitude of the output state.  If this claim is correct, the sampling cost could be an obstacle for any experimental implementation.  However, this conjecture does not apply to QERC, and we show the reason through the sampling analysis below.

To directly evaluate the reconstruction of a distribution, we can employ the Kullback-Leibler divergence (KLD) \cite{mackay2003}.  As we see in Fig.~\ref{Sampling2} (a) in Appendix~\ref{AppendixNumSamp}, the KLD scales polynomially for a fixed number of qubits for all four quantum reservoir models, and the KLD requires a larger sampling number to achieve the same KLD as the number of qubits increases.   This might imply an exponential scaling cost for sampling as the number of qubits increases.  However, the role of the M-layer is not to reconstruct the true amplitude distribution but to provide probability distributions for the linear classifier to classify them.  Fig.~\ref{Sampling} shows the convergence of the accuracy rate as the sampling number increases.  In fact, it converges to the theoretical accuracy rate before the distributions are accurately reconstructed.  Hence, in comparison to the cost of reconstructing the probability amplitude, the necessary sampling number is not high.  The detailed analysis of the sampling cost for the linear classifier is in the Applendix~\ref{AppendixNumSamp}.

The remaining factor for the sampling cost is the scaling of the sampling number to the number of qubits in the quantum reservoir.   For $N=12$, we obtain the same qualitative characteristics shown in Fig.~\ref{Sampling} (see Fig.~\ref{Sampling2} (b-e) in the Appendix~\ref{AppendixNumSamp} ).  More importantly, the case of $N=12$ gives a higher accuracy rate for the same sampling number for all four models, and the accuracy rate converges to the theoretical value at the same sampling number for both numbers of qubits. These strongly indicate that the sampling cost in QERC is much lower than expected by the application of the standard sampling theory.

The analysis of the sampling cost also provides an insight into the differences among the quantum reservoirs.  The ZZ-X model exhibits a different tendency in the small sampling number regime shown in Fig.~\ref{Sampling2}, where it requires a much smaller sampling number than the other models require to achieve the same accuracy (see Appendix~\ref{AppendixNumSamp}).  This can not only be a significant advantage in its implementation. But also be a hint to consider the trade-off between the information scrambling and the clustering of the output state to assist the classical linear classifier.

Information scrambling has to happen during the time evolution of the quantum reservoir.  If this is only the role of the quantum reservoir, the unitary sampled on the Haar measure should give the best result.  Our sampling cost analysis shows that the clustering of the state can happen with certain quantum reservoirs.  In our case, the ZZ-X model achieves both information scrambling and partial clustering of the output states without any programming or optimization of the reservoir parameters.  We investigate the participation ratio in the next section to examine this further.

 \section{Participation Ratio (PR)}

To evaluate the power of reservoir computation, the Lyapunov exponent and memory capacities such as $\tau$-capacity are widely used: the former being a measure of the chaotic dynamics and nonlinearlity in classical reservoirs~\cite{Pathak2017}, while the latter are used in classical and quantum reservoir computing~\cite{Mantas2009, Fujii2017,Nakajima2019,Martinez2021,Han2021,Mujal2023}.  Memory capacity measures how much past information is stored in the reservoir, and unfortunately, such a measure is not appropriate for evaluating the quantum reservoirs used in QERC.  Also, to see what is happening in the computational process through the QERC, we need to know what complexity is generated in the state as well as its degree. Hence, we employ the participation ratio (PR) used in condensed matter physics to measure the local and extended states~\cite{Edwards1972}, such as the Anderson localized and ergodic states, to evaluate the dynamics. For a quantum state $|\Psi\rangle = \sum_l c_l |l\rangle$, the PR is defined as, 
\begin{equation}
\mathrm{PR}\left(|\Psi\rangle\right) = \frac{1}{\sum_l |\langle l |\Psi\rangle|^4} = \frac{1}{\sum_l |c_l|^4} 
\end{equation}
where $|l\rangle$ are the computational basis states for our purpose. Importantly, its inverse, so-called inversed PR, has the same form of a quantity used in the analysis of the neuron dynamics ~\cite{Amari1974}.

The PR of the output state can be highly dependent on the initial state, and so we considered the averaged PR (APR) over the output state set  $\mathbb{D}$ defined by, 
\begin{equation}
{\mathrm{APR}}\left(\mathbb{D}\right) = \frac{1}{\#\,\mathrm{states} }\sum_{|\Psi\rangle \in \mathbb{D}}\left(\frac{1}{\sum_l |\langle l |\Psi\rangle|^4} \right).
\label{eq:apr}
\end{equation}
This value shows how widely the state is spread over the computational basis on average, hence it tells us the representation capability that the quantum reservoir provides. 

Now if the APR is high, an input state would evolve to an output state supported by many computational basis states. Hence, it is more likely that the dynamics yields a strong information scrambling.  However, there is no guarantee that such a set of states is preferable to the classical linear classifier for classification.  From the clustering point of view, all states for a certain class of images should have a distinct characteristic, and hence, APR cannot only describe the property of the quantum reservoir.  The ideal distribution of the output states should have some clustering as well as exploiting the dimension of the Hilbert space.  To see this, we introduce IAPR, which evaluates the overall usage of the dimensionality of the Hilbert space, as
\begin{equation}
{\mathrm{IAPR}}\left(\mathbb{D}\right) = \frac{{(\#\,\mathrm{states})}^2}{{\displaystyle\sum_l \left({\sum_{|\Psi\rangle \in \mathbb{D}} |\langle l |\Psi\rangle|^2}\right)^2}}.
\end{equation}
The IAPR is the APR for the distribution constructed from all the output states as
\begin{equation}
\hat{\rho}_{\mathbb{D}} =\frac{1}{\#\,\mathrm{states}}  \sum_{|\Psi\rangle \in \mathbb{D}} |\Psi\rangle\langle \Psi|.
\end{equation}
If IAPR is large, it indicates that the dimensionality of the Hilbert space is exploited for the set of input states.  Now, we consider the difference, $\Delta\mathrm{PR}=\mathrm{IAPR} -\mathrm{APR}$.  If this difference is large, each state is supported by a relatively small number of basis states, whereas the set of states requires the entire dimension of the Hilbert space.  Hence, this difference can be considered an indicator for clustering the output states.

\begin{figure}[htb]
\centering
\includegraphics[width=0.4\textwidth]{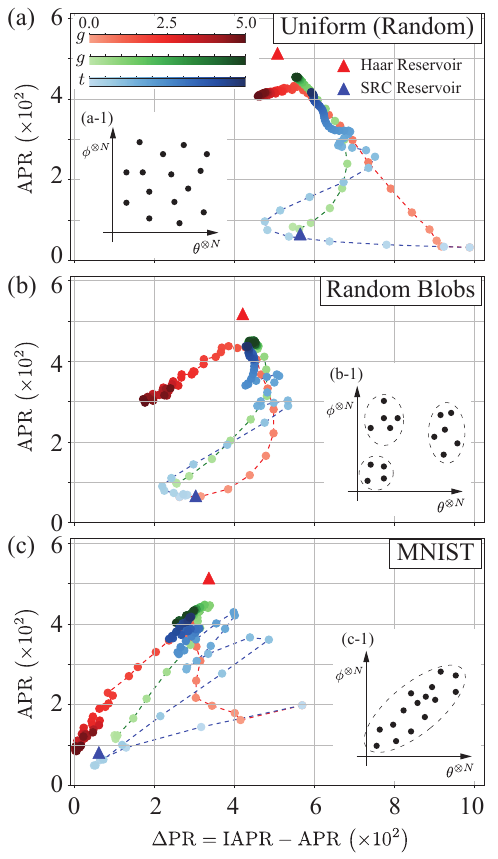}
 \caption{ Plot of the APR vs.~$\Delta \mathrm{PR}$ for ZZX-Ising model using ten qubits. The red and green gradation points show the result with different transverse magnetic strengths $0\leq g/J_0\leq 5$, where we used normal and randomized encoders in red and green, respectively. Here, we set $J_0t=3.5$ with $\alpha = 1.5$.  The blue dots show the changes in APR and IAPR over time. The red and blue triangle markers show the results of the Haar and SRC reservoirs. Now (a) and (b) show the results of sets where $10,000$ initial states were randomly generated. For (a), all states were created by angles $\theta_l$ and $\phi_l$ that are uniformly randomly generated.  Next, (b) illustrates the result for the initial state set generated by isotopic Gaussian blobs, while (c) depicts the results for the MNIST dataset using 60,000 images.}
\label{PR_ZZXIsing}
\end{figure}

Before analyzing the PRs, we need to consider the effect of the bias within the set of input states. These PR values depend on the trend and bias within the set of input states.  To evaluate the quantum reservoirs independently of these trends, we introduce two random input sets.
First, we consider uniformly distributed points in the $2^N$-dimensional space. We generate $10,000$ points randomly and use the same encoder as we did in the benchmark (described in  Ref.~\cite{Akitada2022} and the Appendix~\ref{AppendixEncodeImages}).
However, the completely random input set is not so realistic, as the input data set generally has some bias.  Hence, we also consider a random data set generated by isotropic Gaussian blobs in the high-dimensional square with length $-10$ to $10$ (see the sketch in Fig.~\ref{PR_ZZXIsing} (b-1)). Here we randomly generate $10,000$ points in the high-dimensional space. These points form $10$ isotropic Gaussian blobs (clusters) where the standard deviation of the clusters is $1.2$. These points are mapped to the angles following the encoding process mentioned in Ref.~\cite{Akitada2022} (and Appendix~\ref{AppendixEncodeImages}). We also considered the MNIST data set to evaluate the actual practical case (see (c-1) in Fig.~\ref{PR_ZZXIsing}). 

We show the APR and $\Delta\mathrm{PR}$ values for the ZZ-Ising model with the benchmark against two random reservoirs: the Haar measure sampling and the single rotation random circuit (SRC). These are given by the red and blue triangles in Fig.~\ref{PR_ZZXIsing}, respectively. For the ZZ-Ising situation, the difference between the randomization of the input state appears in the difference between the red (no randomization) and green (randomization) dots.  As the magnetic field strength $g$ increases, both the APR and IAPR decrease, as shown with the red dots. This agrees with the lower accuracy without randomization for the high $g$ regime.  For all three data sets, the ZZ-Ising model gets close to the value given by the Haar sampling, and hence by taking optimal parameter values, the behaviour of this reservoir is expected to be similar to that of Haar sampling.  However, the optimal value for $g$ is not where the APR is the largest. To see this more clearly, we need to investigate the PR values for the ZZ-X model.

\begin{figure*}[thb]
\centering
\includegraphics[width=0.9\textwidth]{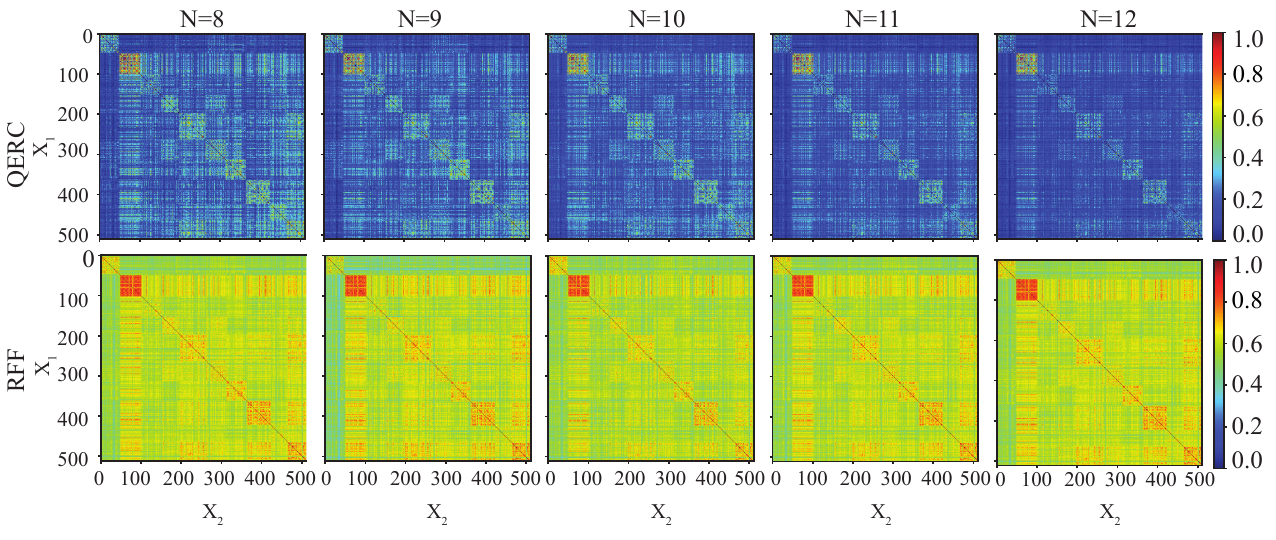}
 \caption{ Normalized kernel functions for QORC and RFF: We randomly selected 512 images ($X_{i=1,2,\ldots,512}$) from the MNIST data set (training set) and computed normalized kernel functions for QORC and RFF. The upper figures are kernel functions for QORC, and the lower figures are kernel functions for RFF. The two axes correspond to the image indices, and the color represents the kernel value. The number of features for both models is $2^N$, where $N$ is from 8 to 12. }
\label{KernelValue}
\end{figure*}

The APR results show that the states are highly spread out in Haar, utilizing a vast number of computational bases. This suggests that distinguishing between the states of different images is challenging. However, considering $\Delta$PR, the PR per image is never the entire space, i.e., $2^N$, approximately half of it. When considering all images collectively, the total space is covered. This implies that images within the same class significantly overlap the computational bases used, whereas images from different classes have fewer common computational bases. To further clarify this point, we examine the kernel matrix, which is widely used in machine learning analysis of feature mappings~\cite{Mikhai2019}. This kernel matrix is the inner product of feature mapping vectors for two inputs, indicating the similarity between data points in feature space. Intuitively, the kernel values are high for images within the same class and lower for those in different classes.

Specifically, we randomly selected 512 images from the training dataset, with each class having approximately 50 images on average. The probability distributions for these images are obtained using Haar-random unitary matrices as a reservoir. Considering our QERC post-processing, we compute the normalized kernel matrix for the outputs processed with Eq.~\ref{eq:PostPross} in the Appendix~\ref{AppendixQERC} and compare it with the classical random feature model, Random Fourier Features (RFF)~\cite{Rahimi2007,Liao_2021}. Now, the normalized kernel matrix is given by,
\begin{equation}
K_{\left(i,j\right)}(x_i,x_j)=\frac{q\left(x_i\right)^T\cdot q\left(x_j\right)}{\sqrt{q\left(x_i\right)^T\cdot q\left(x_i\right)}\sqrt{q\left(x_j\right)^T\cdot q\left(x_j\right)}},
\end{equation}
where $x_{i,j}$ are images and $q_{i,k}$ are corresponding feature map vectors. Note that, in the RFF calculation, we prepared $2^N$ features, where $N$ is the number of qubits we used.

The results are shown in Fig.~\ref{KernelValue}. High similarity is observed between data points within the same class, whereas similarity is significantly lower between different classes. Furthermore, as the number of qubits increases, the similarity between different classes decreases, leading to better class separation. This suggests that our model possesses strong discriminative power in high-dimensional spaces. Additionally, compared with RFF, our model exhibits lower similarity between different classes, enabling more effective discrimination of distinct distributions. At the same time, high similarity is maintained within the same class, ensuring appropriate classification while mitigating overfitting risks.

\begin{figure}[htb]
\centering
\includegraphics[width=0.4\textwidth]{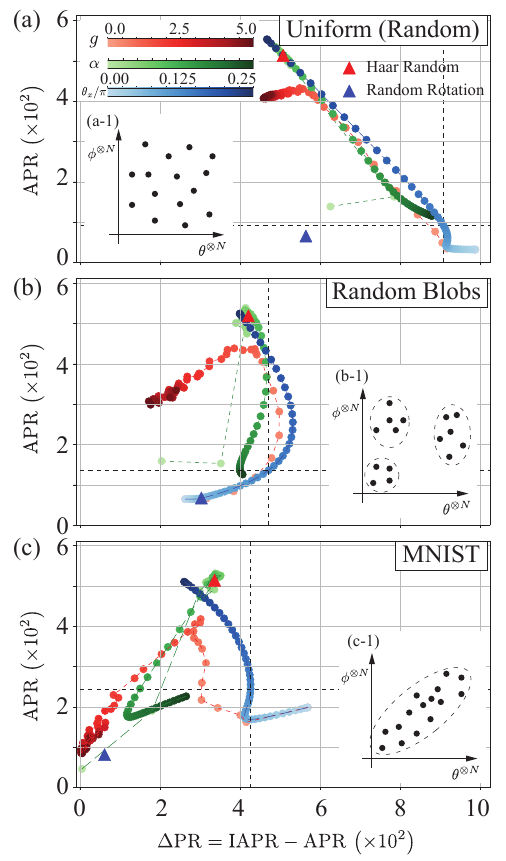}
 \caption{ Plot of the APR vs.~$\Delta \mathrm{PR}$ for the ZZ-X model with ten qubits. The blue line and markers are the results of the ZZ-X model, where we set $0\leq\theta_x \leq0.25\pi/$, $\theta_J =2\pi$, and $\alpha = 1.5$. Here, the red line is the result of the ZZX-Ising model, where we used the same parameters as in Fig.~\ref{PR_ZZXIsing}. The green line results from the XX-Ising model with $g/J_0=0.0$ and $J_0t=3.5$. Black dashed lines indicate the result of $\theta_x = 0.125\pi$ for the ZZ-X model. The red and blue triangle markers show the Haar random and random rotations results. Now (a) and (b) depict the results of the same random initial state sets used in Fig.~\ref{PR_ZZXIsing}  while (c) shows the results from the MNIST dataset.}
\label{PR_ZZX}
\end{figure}

The PR values for the ZZ-X model are shown in Fig.~\ref{PR_ZZX}, with an interesting characteristic being that there are points above the Haar sampling in terms of the APR in the case of the ZZ-X model.  Even more interestingly, the optimal point for the QERC performance does not appear near the Haar sampling.  The blue dots in these figures show the dependency of the PRs on the single-rotation angle.  The optimal point for $\theta_x$ in Fig.~\ref{ZZX} (c) is near $\pi/8$, meaning the optimal point for this model clearly differs from the point given by Haar sampling.  Further the ZZ-X model optimal point is characterized by a lower APR and higher IAPR.  For the cases with a lower APR and a higher IAPR, each output distribution function does not broadly spread on the computational basis. This supports the idea that the ZZ-X model utilizes the Hilbert space as much as other models to support the set of output states, although the distribution of each output state is not so flat.  This clustering effect assists the classical linear classifier distinguish them, reducing the sampling cost.  This PR ratio and the high performance of the ZZ-X model provide evidence that the ZZ-X model realizes the clustering of the output states, meaning that the process to converge to the solution of the task has already started in the time evolution of the quantum reservoir.

\section{Discussion and Conclusion}

The analysis in this paper has shown that Hamiltonian dynamics as simple as ZZ-interaction without the transversal magnetic field can be used to perform non-trivial classical computational tasks.  
In particular, unlike the previous quantum QERC reservoir~\cite{Akitada2022}, no more than two-partite interactions (either physical or effective) are necessary for the QERC to perform. It is well known that the circuit implementation of Haar measure sampling is costly, and hence its circuit depth is long.  The Clifford+T model, however, achieves similar testing accuracy, which is an advantage when the quantum computation is done in a fault-tolerant manner.  

For small scale quantum processing, the Ising models can be implemented in many alternate physical platforms with high accuracy. More importantly, it circumvents the existing criticism for the exponential cost of the sampling measurement due to the fact that the sampling cost does not follow the cost associated with the statistical reconstruction of the distribution function.  In fact, the accuracy rate converges around the fixed sampling number for $N=10$ and $12$. This is an advantage in the implementation.  Additionally, we have confirmed that in other image datasets, such as Fashion-MNIST~\cite{xiao2017} and K-MNIST~\cite{clanuwat2018deep}, we obtained similar results discussed in this paper.

Finally the ZZ-X model outperformed the Haar sampled unitary operation in both the testing accuracy and the sampling cost at the measurement, which directly indicates that the generation of the complexity and the use of it would be two different things in designing quantum machine learning models. 
Our PR analysis shows that the two main processes in quantum algorithms, information scrambling and the convergence to a solution, can be realized by a simple quantum reservoir such as the ZZ-X model without any programming and parameter optimizations.
Our results have shown a new approach to design quantum-classical hybrid machine learning models as well as given an insight into the computational process within quantum reservoirs.

\acknowledgements{\it This work is supported by the MEXT Quantum Leap Flagship Program (MEXT Q-LEAP) under Grant No. JPMXS0118069605, COI-NEXT under Grant No. JPMJPF2221, and the JSPS KAKENHI Grant No. 21H04880.}

\appendix
\section{Quantum Extreme Reservoir Computing (QERC)}
\label{AppendixQERC}
Let us provide a detailed description of Quantum Extreme Reservoir Computation (QERC).  We start with the general framework used in QERC~\cite{Akitada2022} where we consider an input $\mathbf{d}$ being a $D$-dimensional vector. We assume this input to be classical, although QERC can be applied to quantum inputs. Now since $\mathbf{d}$  is classical information, it is necessary to encode it onto the initial quantum state $|\psi(\mathbf{d})\rangle_{\mathrm{I}}$  for the quantum reservoir. Through the quantum reservoir dynamics, the initial state $|\psi(\mathbf{d})\rangle_{\mathrm{I}}$ is mapped to the final state $|\psi(\mathbf{d})\rangle_{\mathrm{F}}$ by the unitary operator $\hat{U}_{\mathrm{QR}}$: $|\psi(\mathbf{d})\rangle_{\mathrm{F}} = \hat{U}_{\mathrm{QR}} |\psi(\mathbf{d})\rangle_{\mathrm{I}}$. 

The final state is then projectively measured into the computational basis, converting the quantum information to classical information.  Here, let us define the notation of the computational basis set for the spin system.  We take the product states of all the eigenstates of $\sigma_l^z$ as the computational basis $|x\rangle$, where $|x\rangle \in \left\{|0\cdots 00\rangle,|0\cdots 01\rangle,\cdots, |1\cdots 11\rangle \right\}$. Because we cannot obtain the amplitude distribution from a single-shot measurement, it is necessary to sample the statistics of the final state. In this model, we use, 
 \begin{equation}
\mathbf{p}(\mathbf{d}\,) 
 = \left(
\begin{matrix}
p_1\\
p_2\\
p_3\\
\vdots \\
p_{2^N}
\end{matrix}
\right)
 = \left(
\begin{matrix}
|\langle 0\cdots 00|\psi_\mathrm{F}(\mathbf{d}\,)\rangle|^2\\
|\langle 0\cdots 01|\psi_\mathrm{F}(\mathbf{d}\,)\rangle|^2\\
|\langle 0\cdots 10|\psi_\mathrm{F}(\mathbf{d}\,)\rangle|^2\\
\vdots \\
|\langle 1\cdots 11|\psi_\mathrm{F}(\mathbf{d}\,)\rangle|^2\\
\end{matrix}
\right),
\end{equation}
for all basis $|x_l\rangle$ with $l=1,\cdots,2^N$. The readout gives us a  $2^N$-dimensional vector.  In our work, the readout is standardized to stabilize the later classical processing: $\mathbf{p}(\mathbf{d}\,) \rightarrow \mathbf{u}(\mathbf{d}\,)$. In more detail, we employ the standardization given by
\begin{equation}
{u}_l = \frac{p_l - \bar{p}}{S},
\label{eq:PostPross}
\end{equation}
where $\bar{p}= (\sum_l^{2^N}p_l)/2^N = 1/2^N$ and $S =  \sum_{l}^{2^N}( p_l -\bar{p})^2$ are the mean value and standard deviation, respectively. We refer to the measurement and the standardization together as the M-layer. Next, $\mathbf{u}$ is sent to the classical classifier where we employ the linear classifier in which the data transfer is given by $\mathbf{y} = f\left(W\mathbf{x}+\mathbf{b}\right)$. Here $\mathbf{y}$ is the final output of the QERC, $W$ and $\mathbf{b}$ are the weight matrix and bias vector. The function $f(\cdot)$ is called an activation function. We employ the softmax activation function~\cite{duchi2011} in the main text because the QERC is applied to the multi-class classification task. Hence, the final output tells the classification class of the input data. 

We use the ``AdaGrad'' classical algorithms \cite{duchi2011} to optimize the weight matrix and bias. It is a gradient descent algorithm that uses landscape information. 
Our interest here is to apply QERC to image classification. As an illustrative example, we use the MNIST dataset for numerical simulations with 60,000 and 10,000 digits images (28$\times$28 pixels) for training and testing, respectively.  The training data prepared in advance has the images $\mathbf{d}^{\mathrm{tr}}$ and labels (answers) $\mathbf{t}^{\mathrm{tr}}$ recorded in an $N_t$-dimensional vector.  In the optimization process, the AdaGrad minimizes the cost function given by
$L = \frac{1}{N_a} \sum_{a=1}^{N_a}\sum_{l=1}^{N_t} t_l^{(a)} \log(y_l^{a})$,
where $N_a$ is the mini-batch size. This cost function is known as cross-entropy~\cite{haykin2009}. 

 \begin{figure}[tb]
\centering
\includegraphics[width=0.45\textwidth]{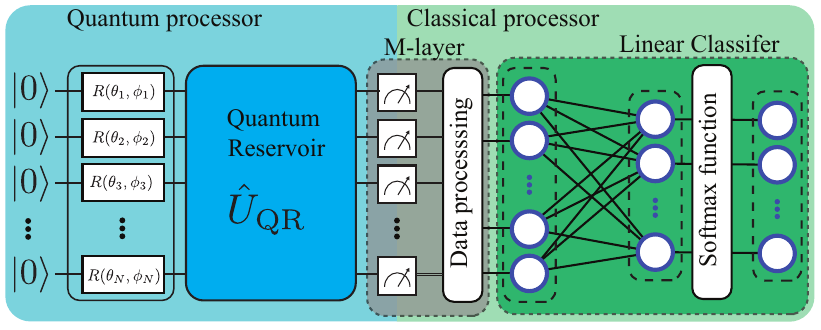}
 \caption{Schematic diagram of the computational process for Quantum Extreme Reservoir Computation (QERC). The bright-blue and bright-green areas represent the quantum and classical processors, respectively. Classical information is encoded into a quantum system using single qubit rotation gates with angles $\theta_l,\phi_l$, where $l=(1,2,\cdots, N)$ and $N$ is the total number of qubits of the quantum processor. In QERC, one can use quantum many-body dynamics (QMD), such as the Ising models discussed in the main text, or a quantum gates sequence as a reservoir. Probability distributions are then determined at the measurement, and data processing is performed (M-layer). Then, the distribution is sent to a linear classifier with the softmax function. The training process is only applied to the classical neural network, while the quantum reservoir is fixed.}
\label{Fig1}
\end{figure}

\subsection{Encode images}
\label{AppendixEncodeImages}
Here, we discuss the sampling cost of QERC to perform it properly. Let us explain the encoder part of QERC. The classical input is encoded into the angles of each qubit. The single qubits can be written as,
\begin{eqnarray}
|\psi_1\rangle= \cos \frac{\theta}{2}|0\rangle + e^{i\phi}\sin\frac{\theta}{2}|1\rangle. 
\label{initialstate}
\end{eqnarray}
where $0\leq \theta < \pi$ and $0\leq \phi < \pi$ are the angles. If $N$-qubits exist, up to $2N$ classical variables can be encoded. However, in the image classification task, such as digit images of MNIST data discussed in the main text, the input data is an image with more information than $2N$ variables. For example, a digit image ($\mathbf{I}$) has $28\time28$ pixels. In other words, it can be assumed that a vector in a $784$-dimensional space: $\mathbf{I} = \sum_{i=1}^{784}c_i \mathbf{e}_i$, where $c_i$ is a pixel value at a basis vector $\mathbf{e}_i$ for $i = 1,2,\cdots,784$. Thus, in Ref.~\cite{Akitada2022}, the Principal Component Analysis (PCA) is used to efficiently decompose images with a new basis set: $\mathbf{I} =  \sum_{j=1}^{784} \tilde{c}_j \mathbf{v}_j$, where the $j$-th element is the $j$-th contribution among the $784$ elements. In the PCA, the new basis vectors are obtained to represent given images efficiently, even using a few basis vectors. It means that there is a priority among the basis. In general, the priority is in the order of $\mathbf{e}_1,\mathbf{e}_2,\cdots$, namely, we refer a coefficient $\tilde{c}_l$ of the $l$-th vector $\mathbf{e}_l$ as $l$-th principal component. Thus, in the QERC, one encodes the images efficiently using the first few components $l=1,2,\cdots,2N$. In more detail, up to $N$-the component are encoded into $\theta$s of qubits, while from $N+1$ to $2N$-th components are encoded into $\phi$s of qubits. To map the components to angles, we employ the conversion  $\tilde{c}_l \to \theta_l $ or $ \phi_l $ given by 
 \begin{equation}
	\theta_{l} \, \text{or} \, \phi_{l} =  \frac{ \pi \left( \tilde{c}_l - \min{ \left[\tilde{c}^{(\text{train)}}_l \right]} \right) }{\max{\left[\tilde{c}^{(\text{train)}}_l \right]} - \min{\left[\tilde{c}^{(\text{train)}}_l \right]}},
\end{equation}
where $\max{[\tilde{c}^{(\text{train)}}_l ]}$ and $\min{[\tilde{c}^{(\text{train)}}_l ]}$ are the maximum and minimum values of $\tilde{c}_l$ across all the training samples respectively. In the MNIST classification,  $\max{[\tilde{c}^{(\text{train)}}_l ]}$ and $\min{[\tilde{c}^{(\text{train)}}_l ]}$ are chosen from $60,000$ images for training. For images not used in training, such as testing images, we note that when $\theta_{l}$ or $\phi_{l}$ goes beyond the range $[0,\pi]$, we truncate the value. 

\section{Sampling cost on QERC}
\subsection{Estimation of Sampling Number}
Let us consider a quantum system with $N$ qubits and run it totally $N_\text{s}$ times to estimate the probability distribution. We measure the system with a computational basis set. In other words, for each run, we can get a bit string $x\in \mathbb{S}_N$, where the set $\mathbb{S}_N$ is a set of all possible N-bit strings: $\mathbb{S}_N = \left\{00\cdots 0,\cdots,11\cdots 1\right\}$.  To simplify the discussion, let us assume that we measure a bit string $x$ with $0\leq N_x \leq N_s$ times. In this assumption, the probability that we get the result is given by
\begin{equation}
P_x(N_x,N_\text{s},p_x) = \frac{N_\text{s}!}{N_x!(N_\text{s}-N_x)}p_x^{N}(1-p_x)^{N-N_x}
\end{equation} 
where $p_x = \text{Tr}{\left(\hat{P}_x\hat{\rho} \right)}$ is an appearance probability of the string $x$, and $\hat{P}_x = |x\rangle \langle x|$ is a projection operator along the computational basis state $|x\rangle$ and $\hat{\rho}$ is a quantum state of $N$-qubits. The expectation value of $N_x$ is given by
\begin{equation}
\langle N_x \rangle = \sum_{N_x = 0}^{N_\text{s}} N_x P_x(N_x,N_\text{s},p_x) = N_sp_x,
\end{equation}
similarly, its variance is given by,
\begin{equation}
\langle \Delta N_x^2 \rangle = \langle N_x^2 \rangle - \langle N_x \rangle^2 = N_\text{s} p_x(1-p_x).
\end{equation} 

Now, let us estimate the probability $p_x$ from the above result. As an estimator for $p_x$, we consider an empirical probability $\tilde{p}_x = N_x/N_\text{s}$. Using the $\langle N_x \rangle$ and $\langle \Delta N_x^2 \rangle$, its expectation is given by
\begin{equation}
\langle \tilde{p}_x \rangle =  \left \langle \frac{N_x}{N_\text{s}} \right\rangle = p_x.
\end{equation}
It indicates that $\tilde{p}_x$ is the unbiased estimator, and its variance is given by
\begin{equation}
\langle \Delta \tilde{p}_x^2 \rangle =  \left \langle\Delta\left( \frac{N_x}{N_\text{s}} \right)^2\right\rangle = {\frac{p_x(1-p_x)}{N_\text{s}}}.
\end{equation}
Thus, its error can be characterized by
\begin{equation}
\epsilon_{\tilde{p}_x} = \sqrt{\langle \Delta \tilde{p}_x^2 \rangle} = \sqrt{\frac{p_x(1-p_x)}{N_\text{s}}}.
\end{equation}
As you can see, the error $\epsilon_{\tilde{p}_x}$ is dependent on $p_x$. As an example, let us consider the two extreme cases. First, let us consider that the probability of the single-bit string $x$ is precisely one. In this case, the error is $\epsilon_{\text{L}} = 0$ . Next, consider a case where the probability distribution is uniform: $p_x = {1}/{2^N}$ for arbitral bit string $x$. Its error can be $\sim 1/\sqrt{2^N N_s}$ for the large $N$. It becomes small as the number of samples $N_s$ increases. As a state is localized, its estimation error gets smaller. While a state is ergodic, its error gets larger. 

 \begin{figure}[htb]
\centering
\includegraphics[width=0.4\textwidth]{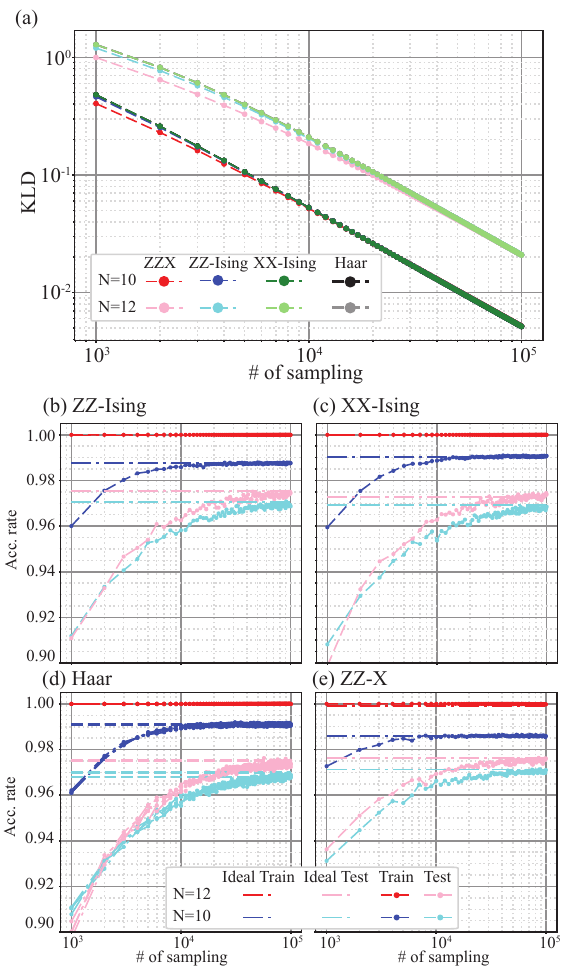}
 \caption{
Plots show the sampling cost of QERC with different quantum reservoirs:  (a) shows the sampling number dependence on Kullback–Leibler divergence (KLD) for different models with $N=10$ and $N=12$. (b)-(e) shows sampling number dependence on the performance with different reservoirs (ZZ-Ising, XX-Ising, the Haar random and ZZ-X model), where we set $g/J_0 =1.0$ for the ZZ-Ising, $g/J_0=0.0$. The color lines show the accuracy rate for the probability distribution by assuming one can get the ideal one. (b) to (e) show the sampling number dependence on the performance for the different reservoirs with $N=10$ and $N=12$. }
\label{Sampling2}
\end{figure}

It is natural to expect that the stochastic fluctuation of empirical probability may affect the performance of the QERC since the distribution is not the same as the ideal distribution. Thus, here let us see a lower bound of the sampling number $N_s$ using $\epsilon_{\tilde{p}_x}$ for QRCs to get good performance. To estimate it, let us assume a constant factor $C$, the minimum fluctuation of the probability distribution that affects the classical neural network. Then we get an inequality of $\epsilon_{\tilde{p}_x}$ as follows, 
\begin{equation}
\epsilon_{\tilde{p}_{x\in \mathbf{S}_N}} < C.
\end{equation}
In general, finding the proper value of the constant $C$ is difficult, but we can estimate it as follows. After the measurement, the probability distribution is renormalized to have an average zero to a variance one, as follows, 
\begin{equation}
{q}_{x\in \mathbb{S}_N} = \frac{p_{x\in \mathbb{S}_N} - \bar{p}}{\sigma},
\end{equation}
where $\bar{p} =\sum_{x\in\mathbb{S}} p_x$ is mean value, and $\sigma = \sum_{x\in\mathbb{S}}( p_x -\bar{p})^2$ is variance of $p_x$. It can be written as $\sigma = (\sum_{x\in\mathbb{S}}p_x^2)-2^{-2N}$. For large systems, because, $2^{-2N}\ll 2^{-N} \leq \sum_{x\in\mathbb{S}}p_x^2$, we assume that $\sigma\approx1/\mathrm{PR}$. Now, in the new parameter space of  $p_x$, its error can be written as,
\begin{equation}
\epsilon_{q_x} = \mathrm{PR} \cdot {\epsilon_{p_x}}.
\end{equation}

Now, it is natural to assume that when the above error is much smaller than that of the distribution of $q_x$ generated by data, QERC works appropriately. Thus, we assume that the stochastic fluctuation is much smaller than the variance of the normalized distribution, which is one. It can be written as $D\cdot \mathrm{PR}\cdot \epsilon_{{p}_x}   < 1$, where $D\gg1$ is a constant that relates to the sensitivity of the classical neural network. Finally, consider a case where $\epsilon_{p_x} \sim 1/\sqrt{2^NN_s}$ is the worst case, and the number of shots can be given ${D^2}\cdot\mathrm{PR}^2/{2^N} < N_s$.  This indicates that the sampling cost is lower as the PR is smaller. 

For example, when a state is uniformly distributed in the Hilbert space, $ \mathrm{PR}\approx 2^N$. Thus, the number of the sampling can be estimated as follows,
\begin{equation}
{D^2}\cdot {2^N} < N_s
\end{equation}
meaning that when $D=10$, we need at least $N_s = 100\times2^N$ sampling to perform QERCs properly.

 \begin{figure}[htb]
\centering
\includegraphics[width=0.4\textwidth]{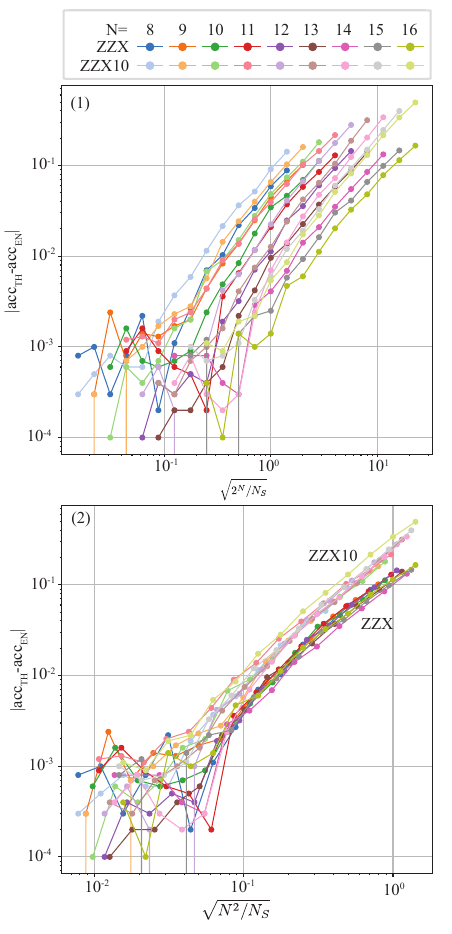}
 \caption{
 To investigate the number of samples required by QERC scales, we evaluated the absolute difference between the test performance using the theoretical probability distribution and the test performance using the empirical distribution (constructed from shots). (1) shows the absolute difference between the rescaled axis of the number of samples and the Hilbert space dimension, giving $\sqrt{2^N/N_s}$, where $N_s$ is the number of samples. (2) shows the case where the rescaled axis is  $\sqrt{N^2/N_s}$. }
\label{Sampling3}
\end{figure}

\subsection{Numerical results with different systems} 
\label{AppendixNumSamp}
First, to see the reconstruction performance of the sampling distribution, we calculate the Kullback-Leibler divergence (KLD) between a theoretical distribution and an empirical distribution, which is numerically obtained by random sampling. The results of KLD with different sizes are shown in Fig.~\ref{Sampling2} (a) .It means that as system size increases, one needs more sampling to achieve the same level, as is well known. 

Next, Figs.~\ref{Sampling2} (b) to (e) show the difference in the required number of samples for each quantum reservoir with respect to the number of qubits.  There would be a trade-off between the advantages (the larger Hilbert space and the larger number of PCA components) and the disadvantage (the statistical overhead of sampling) for the larger number of qubits.  The quantum reservoir with a larger number of qubits would perform better at least at up a certain size in theory due to the larger number of the PCA components can be input and the calculation space is larger, however, due to the reconstruction of the amplitude distribution, it may require a larger number of sampling.  The latter can be seen in the small number sampling regime in Figs.~\ref{Sampling2} (c) and (d).  This indicates that the difference between the amplitude distributions to be distinguished is more detailed with these quantum reservoirs.  By contrast, the ZZ-X model does not have the region where the smaller quantum reservoir achieves a higher accuracy.  This is evidence that the sampling in QERC does not necessarily follow statistical scaling, which is a significant advantage for the implementation. 

To better understand the sample size needed in QERC, we investigated two types of reservoirs: a single ZZX circuit (simply referred to as ZZX) and a model having ten ZZX circuits (denoted as ZZX10). The number of qubits varied from 8 to 16. The first circuit corresponds to the ZZX model described in the main text, where the probability distribution is localized in contrast to other models like the Haar reservoir. The second circuit, ZZX10, applies the ZZX circuit ten times and structurally resembles the discrete-time crystal model introduced in Ref.~\cite{Akitada2022}. Given that ZZX10 alternates between interactions and rotations, its output probability distribution is expected to be adequately spread out. Indeed, for all qubit numbers, the participation ratio (PR) and Shannon entropy of the probability distribution generated by the ZZX10 circuit closely match those of the probability distribution produced by the Haar reservoir. 

The numerical results are shown in Fig.~\ref{Sampling3}. The top panel shows the case where the scaling is $\sqrt{{2^N/N_s}}$. It is clear that the lines corresponding to different qubit numbers do not follow the same curve. This indicates that the required number of qubits does not increase exponentially. While Fig.~\ref{Sampling3} (2) suggests that the necessary number of shots is more likely to be related to $N^2$.

A crucial point is that this scaling law holds regardless of whether the probability distribution is localized or expanded. Additionally, as pointed out in the main text, the ZZX model exhibits a more localized probability distribution than other reservoirs, implying that fewer samples are required than in other models. The common trend for ZZX shows a noticeably smaller slope than that for ZZX10, indicating that the ZZX model requires fewer samples than ZZX10 to achieve similar classification accuracy.

\end{document}